\documentclass[aps,prl,twocolumn,groupedaddress,showpacs,floatfix]{revtex4}
\usepackage{graphicx}

\usepackage{amssymb}

\begin{document}
\title{Measurement of Nuclear Transparency for
the  $A(e,e^\prime \pi^+)$ Reaction}
\author{B.~Clasie$^{1}$}
\author{X.~Qian$^{2}$}
\author{J.~Arrington$^{3}$}
\author{R.~Asaturyan$^{4}$}
\author{F.~Benmokhtar$^{5}$}
\author{W.~Boeglin$^{6}$}
\author{P.~Bosted$^{7}$}
\author{A.~Bruell$^{7}$}
\author{M.~E.~Christy$^{8}$}
\author{E.~Chudakov$^{7}$}
\author{W.~Cosyn$^{9}$}
\author{M.~M.~Dalton$^{10}$}
\author{A.~Daniel$^{11}$}
\author{D.~Day$^{12}$}
\author{D.~Dutta$^{13,2}$}
\author{L.~El~Fassi$^{3}$}
\author{R.~Ent$^{7}$}
\author{H.~C.~Fenker$^{7}$}
\author{J.~Ferrer$^{14}$}
\author{N.~Fomin$^{12}$}
\author{H.~Gao$^{1,2}$}
\author{K.~Garrow$^{15}$}
\author{D.~Gaskell$^{7}$}
\author{C.~Gray$^{10}$}
\author{T.~Horn$^{5,7}$}
\author{G.~M.~Huber$^{16}$}
\author{M.~K.~Jones$^{7}$}
\author{N.~Kalantarians$^{11}$}
\author{C.~E.~Keppel$^{7,8}$}
\author{K.~Kramer$^{2}$}
\author{A.~Larson$^{17}$}
\author{Y.~Li$^{11}$}
\author{Y.~Liang$^{18}$}  
\author{A.~F.~Lung$^{7}$}
\author{S.~Malace$^{8}$}
\author{P.~Markowitz$^{6}$}
\author{A.~Matsumura$^{19}$}
\author{D.~G.~Meekins$^{7}$}
\author{T.~Mertens$^{20}$}
\author{G.~A.~Miller$^{17}$}
\author{T.~Miyoshi$^{11}$}
\author{H.~Mkrtchyan$^{4}$}
\author{R.~Monson$^{21}$}
\author{T.~Navasardyan$^{4}$}
\author{G.~Niculescu$^{14}$}
\author{I.~Niculescu$^{14}$}
\author{Y.~Okayasu$^{19}$}
\author{A.~K.~Opper$^{18}$}
\author{C.~Perdrisat$^{22}$}
\author{V.~Punjabi$^{23}$}
\author{A.~W.~Rauf$^{24}$}
\author{V.~M.~Rodriquez$^{11}$}
\author{D.~Rohe$^{20}$}
\author{J.~Ryckebusch$^{9}$}
\author{J.~Seely$^{1}$}
\author{E.~Segbefia$^{8}$}
\author{G.~R.~Smith$^{7}$}
\author{M.~Strikman$^{25}$}
\author{M.~Sumihama$^{19}$}
\author{V.~Tadevosyan$^{4}$}
\author{L.~Tang$^{7,8}$}
\author{V.~Tvaskis$^{7,8}$}
\author{A.~Villano$^{26}$}
\author{W.~F.~Vulcan$^{7}$}
\author{F.~R.~Wesselmann$^{23}$}
\author{S.~A.~Wood$^{7}$}
\author{L.~Yuan$^{8}$}
\author{X.~C.~Zheng$^{3}$}

\affiliation{$^{1}$Laboratory for Nuclear Science, Massachusetts Institute of Technology, 
  Cambridge, MA, USA}
\affiliation{$^{2}$Triangle Universities Nuclear Laboratory, Duke University, Durham, 
  NC, USA}
\affiliation{$^{3}$Argonne National Laboratory, Argonne, IL, USA}
\affiliation{$^{4}$Yerevan Physics Institute, Yerevan, Armenia}
\affiliation{$^{5}$University of Maryland, College Park, MD, USA}
\affiliation{$^{6}$Florida International University, Miami, FL, USA}
\affiliation{$^{7}$Thomas Jefferson National Laboratory, Newport News, VA, USA}
\affiliation{$^{8}$Hampton University, Hampton, VA, USA}
\affiliation{$^{9}$Ghent University,Ghent, Belgium}
\affiliation{$^{10}$University of the Witwatersrand, Johannesburg, South Africa}
\affiliation{$^{11}$University of Houston, Houston, TX, USA}
\affiliation{$^{12}$University of Virginia, Charlottesville, VA, USA}
\affiliation{$^{13}$Mississippi State University, Mississippi State, MS, USA}
\affiliation{$^{14}$James Madison University, Harrisonburg, VA, USA}
\affiliation{$^{15}$TRIUMF, Vancouver, British Columbia, Canada}
\affiliation{$^{16}$University of Regina, Regina, Saskatchewan, Canada}
\affiliation{$^{17}$University of Washington, Seattle, WA, USA},
\affiliation{$^{18}$Ohio University, Athens, OH, USA}
\affiliation{$^{19}$Tohoku University, Sendai, Japan}
\affiliation{$^{20}$Basel University, Basel, Switzerland}
\affiliation{$^{21}$Central Michigan University, Mount Pleasant, MI, USA}
\affiliation{$^{22}$College of William and Mary, Williamsburg, VA, USA}
\affiliation{$^{23}$Norfolk State University, Norfolk, VA, USA}
\affiliation{$^{24}$University of Manitoba, Winnipeg, Manitoba, Canada}
\affiliation{$^{25}$Pennsylvania State University, University Park, PA, USA} 
\affiliation{$^{26}$Rensselaer Polytechnic Institute, Troy, NY, USA}



\begin{abstract}
We have measured the nuclear transparency of the $A(e,e'\pi^+)$ process 
in $^{2}$H,$^{12}$C, $^{27}$Al, $^{63}$Cu and $^{197}$Au targets. These measurements were performed at the Jefferson Laboratory over a four momentum transfer squared range $Q^2 = 1.1~ \mathrm{to} ~4.7~ (\mathrm{GeV/c})^2$.  The nuclear transparency was 
extracted as the super-ratio of $(\sigma_A/\sigma_H)$ from data to a model of pion-electroproduction from nuclei without $\pi^{_\_}N$ final state interactions. The $Q^2$ and atomic number dependence of the nuclear transparency both show deviations from traditional nuclear physics expectations, and are consistent with calculations that include the quantum chromodynamical phenomenon of color transparency.  
\end{abstract}


\maketitle

In the context of perturbative Quantum Chromo Dynamics (QCD), Brodsky and Mueller~\cite{Mueller:1982} predicted that at sufficiently high momentum transfers, the quark-gluon wave packets of hadrons can be produced as a `color neutral' object of a reduced transverse size. If this compact size is maintained for some distance in traversing the nuclear medium, it would pass undisturbed through the nuclear medium. This is the so-called phenomenon of color transparency (CT). Nuclear transparency, defined as the ratio of the cross section per nucleon 
for a process on a bound nucleon in the nucleus to that from a free nucleon, is the observable used to search for CT. 
A clear signature for the onset of CT would involve a rise in the 
nuclear transparency as a function of $Q^2$. Later works 
\cite{CTreview} have indicated that this phenomenon also occurs 
in a wide variety of models which feature non-perturbative reaction
mechanisms. 

More recently, CT has been discussed in the context of a QCD factorization
theorem derived for meson electroproduction~\cite{factor1}, which states that the meson production amplitude can be expressed in terms of a hard scattering process, a distribution amplitude for the final state meson and a parametrization of the non-perturbative physics inside the nucleon known as Generalized Parton Distributions (GPDs)~\cite{gpd1}. Factorization is expected to be valid for $Q^2 \ge$ 10 (GeV/c)$^2$, however, under certain conditions it may also be applicable at lower $Q^2$~\cite{factor3}. It has been suggested that the onset of CT is a necessary (but not sufficient) condition for factorization to occur~\cite{strikman}. The underlying assumption is that in exclusive ``quasielastic'' production,  the hadron is produced at small inter-quark distances. Thus, the unambiguous observation of the onset of CT is critical as a precondition to the validity of factorization in meson production, and because it would open a new window to study the strong interaction in nuclei.

A number of searches for CT have been carried out in experiments using the $A(p,2p)$, $A(e,e'p)$ reactions, coherent and incoherent meson production from nuclei and pion photoproduction reactions~\cite{ct6}~--~\cite{gammapi}. At high energies the di-jet experiment at Fermi Lab~\cite{fermipi} and $\rho^0$ production at DESY~\cite{hermesrho} are consistent with CT, and it is necessary to include CT to understand shadowing in nuclear deep inelastic scattering~\cite{ff94}. However, the predicted onset of CT at intermediate energies, typical of experiments at JLab and BNL, has not yet been unambiguously observed.  The most recent experiment at JLab to look for CT in $qqq$ hadrons using the $A(e,e'p)$ reaction~\cite{hallc_ct}, does not show any increase of the nuclear transparency up to $Q^2$ = 8.1 (GeV/c)$^2$ and rules out several models predicting an early, rapid onset of CT. One should expect an earlier onset of CT for meson production than for proton scattering~\cite{blattel93}, as it is much more probable to produce a small transverse size in a $q\bar{q}$ system than in a $qqq$ system. Moreover, the 
evolution distances are easily longer than the nuclear radius~\cite{CTreview}, even at moderate $Q^2$, which increases the chances for the small transverse size object to pass undisturbed through the nucleus. These arguments seem to be supported by a recent pion-photoproduction experiment at JLab~\cite{gammapi}. 

We report the first measurement of the nucleon number, $A$, and $Q^2$ dependence of nuclear transparency for the $A(e,e'\pi^+)$ process. The measurement was performed on $^{2}$H, $^{12}$C, $^{27}$Al, $^{63}$Cu and $^{197}$Au nuclei, over a  $Q^2$ range of 1.1 $\mathrm{to}$ 4.7 $(\mathrm{GeV/c})^2$. Measurement of both the $A$ and $Q^2$ dependence of the nuclear transparency is crucial to distinguish between CT-like effects and other reaction-mechanism based energy dependence of the transparency. In this measurement the coherence length for pion production (distance over which the virtual photon fluctuates into a $q\bar{q}$ pair) was smaller than the nucleon radius and was essentially constant, ranging from 0.2 - 0.5 fm over the kinematic range of the experiment.   


The experiment was performed in Hall-C at JLab over the summer and fall of 2004.  A continuous wave electron beam with
energies between 4 and 5.8~GeV and currents between 10 and 80 $\mu$A was incident on solid foil targets of $^{12}$C, $^{27}$Al, $^{63}$Cu and $^{197}$Au, and 4~cm long  liquid hydrogen and deuterium targets. The $^{27}$Al target was used to mimic the cell walls of the liquid target, and enabled the events from the walls to be subtracted from the hydrogen and deuterium data. The scattered electrons and the electroproduced pions ($\pi^+$) were detected in coincidence using the Short Orbit Spectrometer (SOS) and the High Momentum Spectrometer (HMS) respectively. The kinematic settings of the measurements are shown in Table ~\ref{table1}. The center of mass energy of the hadron system, $W$, for all these settings was above 2.14 GeV in order to avoid the resonance region. In addition to the kinematics shown in Table~\ref{table1}, data were collected using the hydrogen target over a range of $\pm$ 5$^o$ around the momentum transfer direction, $\theta_{q} = \theta_{\pi}$, for each $Q^2$ setting (except when limited by the hardware constraint $\theta_{\rm HMS}>$ 10.6$^{\circ}$). These data helped 
develop a model of the elementary pion electroproduction cross section covering a greater range in $\theta_q$. 
For two $Q^2$ settings (2.15 and 3.91 (GeV/c)$^2$) data were also collected at a larger electron scattering angle, in order to perform a Rosenbluth separation as a check of the reaction mechanism. The results of the Rosenbluth separation will be presented in a forthcoming article. 
\begin{table}
\caption{\label{table1} The central kinematics of the experiment.}
\begin{tabular}{|c|c|c|c|c|c|c|c|}\hline
 $Q^2$ & $-t$ & $E_e$ & $\theta_{e^\prime}^{\rm SOS}$ &$E_{e^\prime}$ &
 $\theta_{\pi}$ & $\theta_{\rm HMS}$ & $p_\pi$ \\
 (GeV/c)$^2$ & (GeV/c)$^2$ & GeV & deg & GeV & deg & deg & GeV/c \\\hline
1.10 &0.050 &4.021 &27.76 &1.190 &10.58 &10.61 &2.793 \\
2.15 &0.158 &5.012 &28.85 &1.730 &13.44 &13.44 &3.187 \\
3.00 &0.289 &5.012 &37.77 &1.430 &12.74 &12.74 &3.418 \\
3.91 &0.413 &5.767 &40.38 &1.423 &11.53 &11.53 &4.077 \\
4.69 &0.527 &5.767 &52.67 &1.034 & 9.09 &10.63 &4.412 \\\hline      
\end{tabular}
\vspace{-0.5cm}
\end{table}

The SOS gas \v{C}erenkov counter was used to select the scattered electrons with an efficiency of better than 99.2\%. The pions were selected using the HMS  aerogel~\cite{aerogel} and gas \v{C}erenkov counters, with better than 98.8\% and 99.1\% efficiency, respectively. The HMS aerogel \v{C}erenkov counter was used to select pions only for HMS central momenta $<$~3.2~(GeV/c), because they were below the pion threshold in the HMS gas \v{C}erenkov counter. At higher momentum settings, the HMS gas \v{C}erenkov counter alone was sufficient for selecting pions. The spectrometer acceptance was determined with a relative uncertainty of 1\% between targets using a Monte Carlo simulation of the experimental apparatus, as described below. For each run, the $e-\pi^+$ coincidence events were corrected for random coincidences. 
The charge weighted coincidence yield was also corrected for blocked coincidences ($< $ 0.7\%), loss of synchronization between detectors ($< $ 1.0\%), trigger inefficiency ( $< $ 0.5\%), electronic dead time ( $< $ 1.0\%), computer dead time ( $< $ 25\%, known to much better than 1\%), tracking inefficiency ( $< $ 4.0\%) and particle absorption in the spectrometer material (5.0\% known to better than 0.5\%).  Events from multiple-pion production were rejected with a cut on the missing mass spectrum for each target. Using a simulation of the multiple-pion background it was estimated that the contamination from such events was $<$ 0.4\%. 

The standard Hall C Monte Carlo simulation code SIMC was used to simulate the experimental apparatus~\cite{ben_thesis}. 
The $p(e,e'\pi^+)n$ cross section needed in the model was iterated until 
there was good agreement between the simulation and the experimental data. The iteration was performed separately for each of the kinematic settings in Table ~\ref{table1}. A parametrization of the $p(e,e'\pi^+)n$ cross section from previous data~\cite{tanja_thesis} was used as the 
starting model. 

\begin{figure}
\begin{center}
\includegraphics[width=75mm, height=70mm]{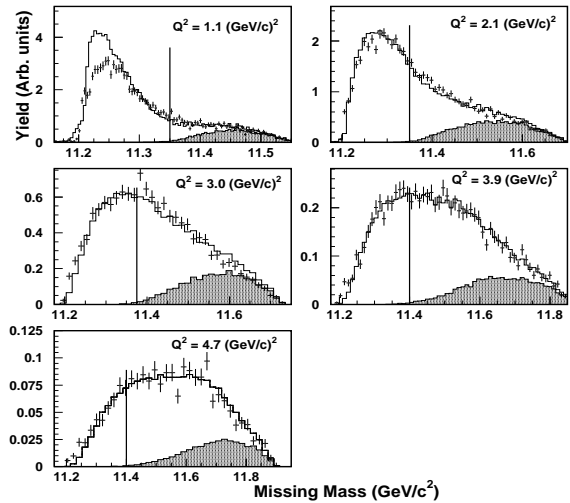}
\caption[]{The missing-mass spectra for $^{12}C(e,e'\pi)$; the crosses are data, the solid line is the simulation which is normalized to the data. The shaded region shows the simulated multi-pion background. The vertical line indicates the position of the multi-pion cut as defined in the text. }
\label{eepi_dist}
\vspace{-1.0cm}
\end{center}
\end{figure}

The Fermi motion of the nucleons in $A>1$ targets was simulated by folding the elementary cross section with a spectral function for the target. For each target, an appropriate Independent Particle Shell Model (IPSM) spectral function was used~\cite{013longpaper}. 
The simulation includes several corrections such as pion decay within the spectrometer, external and internal bremsstrahlung radiation and pions punching through the collimators at the spectrometer entrance. It also included corrections due to various reaction mechanism of the $A(e,e'\pi^+)$ process such as the Coulomb distortion of the incoming and scattered electrons, the bound proton in the target nuclei being off-shell and Pauli exclusion of the recoiling neutron. 

The phase space for multiple-pion production within the spectrometer acceptance was simulated assuming a quasi-free single pion production and a uniform phase space distribution of the additional pions. The simulated multi-pion strength was normalized to the tail of the experimental missing mass spectra. The multiple-pion simulation was used to determine the location of the cut 
on the experimental missing mass spectrum such that the contamination from 
multiple-pion events was less than 0.4\%. This allowed the missing mass cut to be 
placed $\sim$ 10 -- 50 MeV/c$^2$ above the actual kinematic threshold for two-pion production. The simulation 
was able to reproduce the shapes of the measured $W$, $Q^2$ and $|t|$ distributions versus the missing mass reasonably well for all targets and $Q^2$ settings.  Representative missing-mass spectra for $^{12}C(e,e'\pi)$ are shown
together with the Plane Wave Impulse Approximation (PWIA) simulation in Fig.~\ref{eepi_dist} for
all $Q^2$ settings. 
The agreement between the missing mass spectra obtained from data and 
simulation improve with increasing $Q^2$. The discrepancy seen at $Q^2$ = 1.1 (GeV/c)$^2$ can be attributed to the reaction mechanisms missing from the simulations such as final state interactions between the knocked-out neutron and the residual nucleons (nN-FSI) and short range correlations. 

In order to extract the nuclear transparency from the experimental yields, 
the cross section for the bound proton must be corrected for the effects of Fermi motion, Pauli blocking, the off-shell properties of the proton and the acceptances of the spectrometers. In order to account for these 
effects, the nuclear transparency was formed using the experimental charge normalized yield, $\bar Y$, divided by the charge normalized Monte Carlo equivalent yield, $\bar Y_{\rm MC}$. For a given target,
with nucleon number, $A$, the nuclear transparency was defined as:
\begin{equation} \label{equ:nucltransp}
T = 
{\left( \bar Y/ \bar Y_{\rm MC} \right)_A} /
{\left( \bar Y/ \bar Y_{\rm MC} \right)_{\rm H}},
\end{equation}
where the denominator is the ratio of the yields from the $^1$H target.
As the Monte Carlo simulation does not include final-state interactions between the pion and the residual nucleons, the nuclear transparency is a measure of these final-state interactions, and the reduction of these interactions is a signature of CT.

\begin{figure}
\begin{center}
\includegraphics[width=75mm,height=55mm]{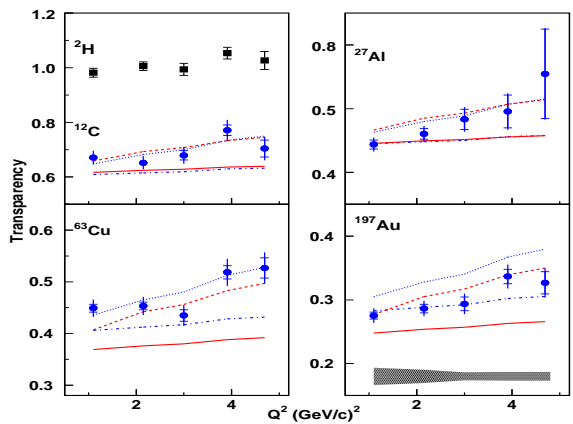}
\caption
{Nuclear transparency, T, vs. $Q^2$ for $^2$H and $^{12}$C (left, top panel), $^{27}$Al (right, top), $^{63}$Cu (left, bottom) and  $^{197}$Au (right, bottom). The inner error bars are the statistical uncertainties and the outer error
bars are the statistical and point-to-point systematic uncertainties added in
quadrature. The dark band in the bottom right panel is the $Q^2$ dependent model uncertainty, and is the same for all nuclei. The solid and dashed lines are Glauber and Glauber plus CT calculations, respectively~\cite{Larson06ge}. 
Similarly, the dot-dash and dotted lines are Glauber and Glauber plus CT calculations, respectively~\cite{wim}. These calculations also include the effect of short range correlations (SRC). 
}  

\label{fig1}
\vspace{-1.0cm}
\end{center}
\end{figure}

Traditional nuclear physics calculations based on the Glauber multiple
scattering mechanism~\cite{glauber} are expected to be energy-independent ( because the $\pi$-nucleon cross section is constant for the energies in this experiment).  To investigate the energy dependence, the extracted 
nuclear transparency is shown as a function of $Q^2$ in
Fig.~\ref{fig1}. The point-to-point ($Q^2$ dependent) systematic uncertainty is 2.4 -- 3.2\%, dominated by uncertainty in the spectral function (1\%) and the iteration procedure (1\%). There is an additional normalization systematic uncertainty of 1.1\% (not shown in the figure) with pion absorption correction (0.5\%), and target thickness (1\%) being the main sources. The $Q^2$ dependent model uncertainty is 7.6\%, 5.7\%, 3.5\%, 3.8\% and 3.8\% for $Q^2$ = 1.1, 2.1, 3.0, 3.9 and 4.7 (GeV/c)$^2$ respectively. This uncertainty was determined from the change in $Q^2$ dependence of the transparency when using two different spectral functions and two different Fermi distributions in the simulation, and the $Q^2$ dependent uncertainty from reactions mechanisms not included in the simulation (estimated by quantifying the difference in shape of the missing-mass spectra from data and simulation) added in quadrature. The $Q^2$ dependent model uncertainty is shown as a dark band in the bottom right panel of Fig.~\ref{fig1}. There is an additional 7.0\% normalization type model uncertainty, independent of $Q^2$, not shown in the figure.
The observed $Q^2$ dependence of the transparency deviates from the 
calculations without CT of Larson {\it et al.} and Cosyn {\it et al.}~\cite{Larson06ge,wim}, and are in better 
agreement with the CT calculations of the same authors. Larson {\it et al.} use a semi-classical Glauber multiple scattering approximation, while Cosyn {\it et al.} use a relativistic version of Glauber multiple scattering theory. Both groups incorporate CT using the quantum diffusion model of Ref.~\cite{farrar} with the same parameters $\tau$ = 1~fm/c and $M_h^2$ = 0.7 GeV$^2$.

In addition to the $Q^2$ dependence,
the dependence of the nuclear transparency on $A$,
is important in the search of CT effects and  
is examined by fitting the transparency as a
function of $A$ at fixed $Q^2$ to the form $T = A^{\alpha-1}$. 
The parameter $\alpha$ is found to be $\sim$ 0.76 in fits to the pion-nucleus scattering cross sections~\cite{carrol1}, and it is expected to be energy independent. An energy dependence of the parameter $\alpha$ (which quantifies the A dependence of nuclear transparency) is a signal for CT-like effects. Our results shown in Fig.~3, indicate that the
energy dependence of the parameter $\alpha$ deviates significantly from the conventional nuclear physics expectation. The systematic uncertainties shown include contributions from the fitting error and the model uncertainties. Our results are in reasonable agreement with $\alpha$ extracted from the calculations (with CT) of Larson {\it et al.}~\cite{Larson06ge} but are systematically lower than the calculations (with CT and short range correlations) of Cosyn {\it et al.}~\cite{wim}.   

\begin{figure}
\begin{center}
\includegraphics[width=75mm]{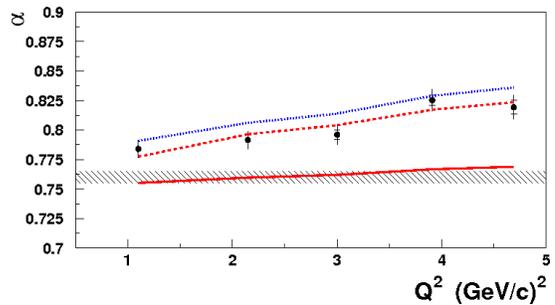}
\caption
{The parameter $\alpha$ (from T = A$^{\alpha-1}$) is shown vs  $Q^2$. The inner error bars are the statistical uncertainty and the outer error bars are the quadrature sum of statistical and systematic and model uncertainties. The hatched band is the value of $\alpha$ extracted from pion-nucleus scattering data~\cite{carrol1}. The solid, dashed, and dotted  lines are $\alpha$ obtained from fitting the A dependence of the theoretical calculations, Glauber, Glauber+CT ~\cite{Larson06ge}, and Glauber+SRC+CT~\cite{wim} respectively. 
}  
\label{fig2}
\end{center}
\vspace{-1.0cm}
\end{figure}

These results seem to confirm the predicted early onset of CT 
in mesons compared to baryons. 
Our results, together with the previous meson transparency measurements~\cite{gammapi, hermesrho}, 
suggest a gradual transition to meson production with small inter-quark separation, and the onset of reaction mechanisms necessary for QCD-factorization at $Q^2$ values of a few (GeV/c)$^2$. These results also put severe constraints on early models of CT which predict a dramatic transition with a threshold-like behavior.     


In summary, we have measured the nuclear transparency of pions from
$Q^2$ = 1.1 to 4.7 (GeV/c)$^2$ over a wide range of $A$ (2 - 197). Both
the energy dependence and the $A$ dependence of the transparency show
deviations from the traditional nuclear physics expectations and are in 
agreement with CT calculations~\cite{Larson06ge,wim}. It is important to extend these measurements to $Q^2 \sim $ 10 (GeV/c)$^2$, where the largest CT effects are predicted, in order to establish the onset of CT effect on a firm footing.

We acknowledge the outstanding support of JLab Hall C technical staff and
Accelerator Division in accomplishing this experiment. This work was supported in part by the U.~S.~Department of Energy, the U.~S.~National Science Foundation, and the Natural Science and Engineering Research Council of Canada.
This work was supported by DOE contract DE-AC05-84ER40150
under which the Southeastern Universities Research Association
(SURA) operates the Thomas Jefferson National Accelerator Facility.

\end{document}